\documentclass[10pt,floats,floatfix,superscriptaddress,nofootinbib]{iopart} 

\expandafter\let\csname equation*\endcsname=\relax 
\expandafter\let\csname endequation*\endcsname=\relax
\usepackage{amsmath}
\usepackage{geometry}

\usepackage{amssymb}
\usepackage[T1]{fontenc}
\usepackage[utf8]{inputenc}
\usepackage{color}
\usepackage{bm}
\usepackage{graphicx}
\usepackage{bbm}
\usepackage{dsfont}
\usepackage[singlespacing]{setspace}

\begin{document}
\title[]{Two-body metal-insulator transitions in the Anderson-Hubbard model}

\author{F Stellin$^1$ and G Orso$^2$}

\address{$^{1,2}\hspace{0.05cm}$Universit\' e de Paris, Laboratoire Mat\' eriaux et Ph\' enom\`enes Quantiques, CNRS, F-75013, Paris, France\\
$^{1}\hspace{0.05cm}$Universit\' e Paris-Saclay, ENS Paris-Saclay, CNRS, Centre Borelli, F-91190, Gif-sur-Yvette, France}

\eads{\mailto{filippo.stellin@univ-paris-diderot.fr}, \mailto{giuliano.orso@univ-paris-diderot.fr}}
\vspace{10pt}

\begin{abstract}
We review our recent results on Anderson localization in systems of two interacting particles coupled by contact interactions. Based on an exact mapping to an effective single-particle problem, we numerically investigate the occurrence of metal-insulator phase transitions for the pair in two- (2D) and three-dimensional (3D) disordered lattices. 
In two dimensions, we find that interactions cause an exponential  enhancement of the pair localization length with respect to its single-particle counterpart, but do not induce a delocalization transition. In particular we show that previous claims of 2D interaction-induced Anderson transitions 
are the results of strong finite-size effects.
 In three dimensions we find that the pair undergoes a metal-insulator transition belonging to the same (orthogonal) universality class of the noninteracting model. We then explore the phase diagram in the space of energy $E$, disorder $W$ and interaction strength $U$, which reveals a rich and counterintuitive structure, endowed with multiple metallic and insulating phases. We point out that this phenomenon originates from the molecular and scattering-like nature of the pair states available at given energy and disorder strength.
\end{abstract}

\section{Introduction}
 The propagation of waves in random media can be suppressed when a certain degree of disorder is exceeded, due to the interference between the multiple scattering paths generated by the impurities. This phenomenon, known as Anderson localization, was initially predicted for spin excitations in semiconductors~\cite{Anderson1958} and has been reported for different kinds of waves, such as light waves~\cite{Segev:LocAnderson2DLight:N07}, ultrasound~\cite{vanTiggelen:Multifrac:PRL09}, microwaves~\cite{Chabanov:StatisticalSignaturesPhotonLoc:N00} and atomic matter waves~\cite{Billy:AndersonBEC1D:N08,Roati:AubryAndreBEC1D:N08,Lopez:ExperimentalTestOfUniversality:PRL12}. Its occurrence is ruled by either spatial dimension and symmetries of the model. For example, when both time-reversal and spin-rotation invariances are preserved, all wave-functions are exponentially localized  in one and two spatial dimensions. In three dimensions, the energy spectrum contains both regions of localized and of extended states, which are separated by
  one or more energy thresholds, dubbed mobility edges~\cite{MottMobilityEdge:AdvPhys67}. 

The presence of interactions can substantially modify the one-particle scenario \cite{FleishmanAnderson:ATInt:PRB79}, as it was confirmed since the earliest studies on interacting electrons in disordered metals~\cite{Giamarchi:ALInt1D1988}, among which the scaling theory proposed by Finkelstein~\cite{Finkelstein:IntElec:1983}. 
The puzzling evidence of a metal-insulator transition observed in two-dimensional silicon metal-oxide-semiconductor field-effect transistors~\cite{Kravchenko:MIT2DExp:PRB94} has also fostered attention on the field, with further theoretical explanations \cite{Punnoose:Science2005} and numerical proofs~\cite{Shepelyansky:electronsMIT2D:2000,Shepelyansky:ElecMIT2DSpectStat:2000}.
In more recent years, the topic of many-body localization, namely the transposition of the concept of Anderson localization to the Fock space for interacting disordered
systems with a finite density of particles, has attracted an increasing interest~\cite{GornyiPRL2005,Altshuler:MetalInsulator:ANP06,Abanin:RMP2019}. While in one-dimensional systems the existence of a many-body mobility edge between non-thermalized states and ergodic ones has been proved for particles subject to short-range interactions, 
its fate in higher-dimensional systems is still unclear~\cite{DeRoeck:PRB2017,WahlNatPhys2019}. 

A complementary viewpoint on interaction-induced delocalization transitions 
 focuses on the transport properties of few-body systems in lattices of infinite size, a problem which is computationally more accessible, especially in dimensions higher than one. 
 This problem was first addressed thirty years ago in  the seminal works of Dorokhov~\cite{Dorokhov:JETP1990} and Shepelyansky~\cite{Shepelyansky:AndLocTIP1D:PRL94}, 
 who showed that, in one dimension, interactions can cause a sizable increase of the pair localization length, with respect to the single-particle one. Since then, the effect  has been thoroughly investigated in the presence of long- \cite{Shepelyansky:electronsMIT2D:2000} and short-range interactions~\cite{vonOppen:AndLocTIPDeloc:PRL96,Frahm1999,Frahm:EigStructAL1DTIP16,Thongjaomayum:PRB2019}. In two dimensions, however, very few numerical studies have been pursued so far~\cite{Ortugno:AndLocTIPDeloc:EPL99,Roemer1999,Chattaraj2018}, as the large computational cost severely limits the system sizes that can be attained. Interestingly, it was claimed~\cite{Ortugno:AndLocTIPDeloc:EPL99,Roemer1999} that interactions can induce a pair delocalization transition, in contrast to previous theoretical
 arguments~\cite{Imry:CohPropTIP:EPL95} suggesting that the localization length of the two-particle system remains always finite.  The exploration of the 3D version of the
 same problem has remained unattempted so far. 
 
In this proceeding article, we review our recent numerical works~\cite{Stellin:PRB2019,Stellin:3DPhaseDiags:2020, Stellin:No2BATs2D:2020} on two-body Anderson localization in two and in three spatial dimensions, by highlighting the most important results.

\section{Model and method}
We consider a system of two particles moving in a disordered lattice and coupled by contact (Hubbard) interactions. The two particles can be
 bosons or fermions in the spin-singlet state, so that the orbital part of the wavefunction is symmetric under particles exchange and
is therefore sensitive to the interactions. 
The Hamiltonian of the two-particle system in the lattice basis $\{|\bm{i},\bm{j}\rangle\}$ is given by:
\begin{equation}
\label{eqn:TIPHam}
\hat{H} = - J\sum_{\{\bm{i},\bm{k}\},\bm{j}} |\bm{i},\bm{j}\rangle\langle \bm{k},\bm{j}| - J\sum_{\bm{i},\{\bm{j},\bm{k}\}}|\bm{i},\bm{j}\rangle\langle \bm{i},\bm{k}| + \sum_{\bm{i},\bm{j}} (v_{\bm{i}}+v_{\bm{j}})
|\bm{i},\bm{j}\rangle \langle \bm{i},\bm{j}|+U\sum_{\bm{i}}|\bm{i},\bm{i}\rangle\langle\bm{i},\bm{i}|\hspace{0.05cm},
\end{equation}
where $J$ is the tunnelling amplitude between nearest-neighbour sites, $U$ is the Hubbard interaction strength  
and $v_{\bm{i}}$ represents the value of the disorder potential at site $\bm{i}$. 
As in Anderson's original work~\cite{Anderson1958}, we assume that the random energies  are uncorrelated, $\overline{v_{\bm{i}}v_{\bm{j}}}=\overline{v_{\bm{i}}^{2}}\delta_{\bm{i},\bm{j}}$ and obey a uniform on-site distribution:
\begin{equation}
\label{eqn:RandPotStat}
P(v_{\pmb{i}})=\frac{1}{W}\Theta\biggl(\frac{W}{2}-\lvert v_{\pmb{i}}\rvert\biggr)\hspace{0.05cm},
\end{equation}
where $\Theta(x)$ denotes the Heaviside unit-step function and $W$ is the disorder strength. In the subspace of orbitally symmetric wave-functions, the stationary 
Schrodinger equation $\hat{H}|\Psi\rangle=E |\Psi\rangle$ can be mapped (exactly) onto a close equation for the diagonal amplitudes 
$\Psi_{\bm{n}}=\langle\bm{n},\bm{n}|\Psi\rangle$ of the wavefunction as in Refs.~\cite{Dufour:PRL2012,vonOppen:AndLocTIPDeloc:PRL96}
 \begin{equation}
\label{eqn:EffKerMod}
\frac{1}{U}\Psi_{\bm{n}}=\sum_{\bm{m}}K_{\bm{n}\bm{m}} \Psi_{\bm{m}}\hspace{0.05cm},
\end{equation}
 where  $K_{\bm{n}\bm{m}}=\langle \bm{n},\bm{n}| (E\hat{\mathbbm{1}}-\hat{H}_{0})^{-1} | \bm{m},\bm{m}\rangle$. Here  $\hat{H}_{0}=\hat{H}(U=0)$ represents the noninteracting 
 two-particle Hamiltonian and $E$ the total energy of the pair.  Equation (\ref{eqn:EffKerMod}) represents an eigenvalue problem and can therefore be interpreted as an 
 effective single-particle Schrodinger equation, describing the centre-of-mass motion of the pair. 
Notice that, in this effective model, the pseudo-energy is given by the inverse of the interaction strength, $\lambda=1/U$, while the total energy $E$ is a continuous parameter of the matrix $K$.
 Unlike the single-particle Anderson model, where the electron can tunnel only between nearest neighboring sites, the matrix $K$ in Eq.~(\ref{eqn:EffKerMod}) contains hopping processes between arbitrarily distant sites.
 
As in standard transfer matrix calculations, we address the localization properties of the effective model, Eq.~(\ref{eqn:EffKerMod}) by computing 
the pair transmission amplitude along bar-shaped $d$-dimensional lattices with $N=M^{d-1}L$ sites and length $L \gg M$. 
In order to reduce finite-size effects, we impose periodic boundary conditions in the transverse directions and open boundary conditions along 
the transmission axis. Labelling as $\{|\phi_r\rangle, \varepsilon_{r}\}$ the eigenbasis of the one-particle Hamiltonian  
\begin{equation}\label{Hsp}
 \hat H^\textrm{sp}=- J\sum_{\{\bm{i},\bm{k}\}}  |\bm{i}\rangle\langle \bm{k}| 
 + \sum_{\bm{i}} v_{\bm{i}} |\bm{i}\rangle \langle \bm{i}|,
 \end{equation}
the matrix elements of the effective model in Eq. \eqref{eqn:EffKerMod} can be written as
 \begin{equation}
\label{eqn:EKerComp}
K_{\bm{m}\bm{n}}=\sum\limits_{r=1}^{N}\phi^{*}_{r,\bm{n}}\phi_{r,\bm{m}}G_{\bm{m},\bm{n}}^{sp}(E-\varepsilon_{r})\hspace{0.05cm}, 
\end{equation}
 where $\phi_{r,\bm n}=\langle \bm n|\phi_r \rangle $ and $G^{\mathrm{sp}}_{\bm{m},\bm{n}}(E)= \sum_{s=1}^{N}\phi_{s,\bm{m}}^{*}\phi_{s,\bm{n}}(E-\varepsilon_{r})^{-1}$ 
 are the entries of the matrix resolvent associated to 
$\hat H^{sp}$. Thanks to the open boundary conditions along the longitudinal direction,  the single-particle Hamiltonian 
possesses a block tridiagonal structure that can be exploited to efficiently compute the Green's function $G^{\mathrm{sp}}$ via matrix inversion
based on a generalized Thomas algorithm~\cite{Jain:2007}. In this way, the total number of elementary operations needed to compute the matrix $K$ 
scales as $M^{4(d-1)}L^{3}$ instead of $M^{4(d-1)}L^{4}$, as expected from standard algorithms for the inversion of a generic symmetric matrix.

The transmission amplitude of the pair between two transverse sections of the bar, the first situated at the head and the second one at a distance $n_z$ inside the bar,
 is defined in terms of the resolvent matrix $G^{\lambda}=(\lambda \mathbbm{1}-K)^{-1}$
as 
\begin{equation}\label{eqn:TransAmpl}
t_{1n_z}=\sum_{\bm{m_{\perp}}, \bm{n_{\perp}}}       J^{2} \lvert G^{\lambda}_{(\bm{m_{\perp}},1), (\bm{n_{\perp}},n_z)}\rvert^{2}\hspace{0.1cm},
\end{equation}
where the sum is taken over all sites $\bm{ m_{\perp}}, \bm{n_{\perp}}$ of two corresponding sections. The logarithm  of the transmission amplitude (\ref{eqn:TransAmpl}) is a self-averaging quantity.
For each disorder realization, we evaluate it at regular intervals along the bar and apply  a linear fit to the data,  $\ln t^{\mathrm{fit}}_{1n_z}=p n_z+q$. For a given value of the interaction strength, we evaluate the Lyapunov exponent $\gamma=\gamma(M,U)$ as $\gamma=-\overline{p}/2$, where $\overline{p}$ is the average of the slope over the different disorder realizations. 
We then infer the transport properties of the system from the behaviour of the reduced localization length, which is defined as  $\Lambda_{M}=(M \gamma)^{-1}$. In the metallic phase $\Lambda_{M}$ increases as $M$ increases, whereas in the insulating phase it does the opposite. At the critical point, $\Lambda_{M}$ becomes constant for values of $M$ sufficiently large. Hence the critical point $U=U_c$ of the Anderson transition can be identified by plotting the reduced localization length versus $U$ for different values of the transverse size $M$ and looking at their common crossing points.
In the following sections we will consider the hopping amplitude as the energy scale of the system and thus we will set $J=1$.\\

\section{Absence of 2D MITs for the pair}

In two-dimensional lattices we have computed the reduced localization length in the region of the parameters space in which a delocalization transition was first predicted~\cite{Ortugno:AndLocTIPDeloc:EPL99,Roemer1999}. In Ref.~\cite{Ortugno:AndLocTIPDeloc:EPL99}, the metal-insulator transition was predicted to occur for $E=0$ and $W=9.3\pm0.5$, based on transmission-amplitude calculations performed on rectangular strips of width $M=10$ and length $L=62$ at most. \\
\begin{figure} 
           \label{fig:Lambda2d}
 	\includegraphics[width=0.485\textwidth]{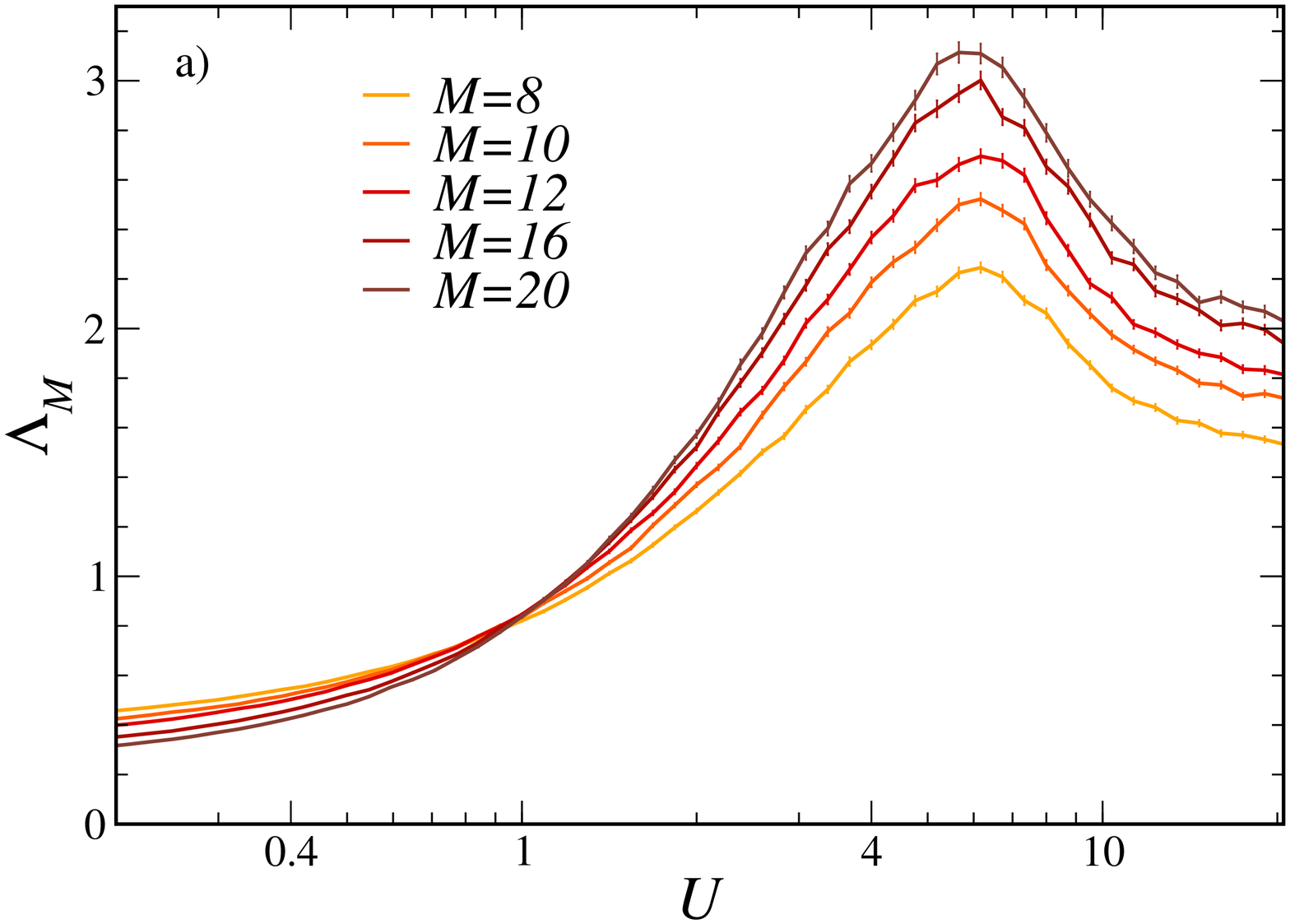}
 	\hspace{\fill}
 	\includegraphics[width=0.485\textwidth]{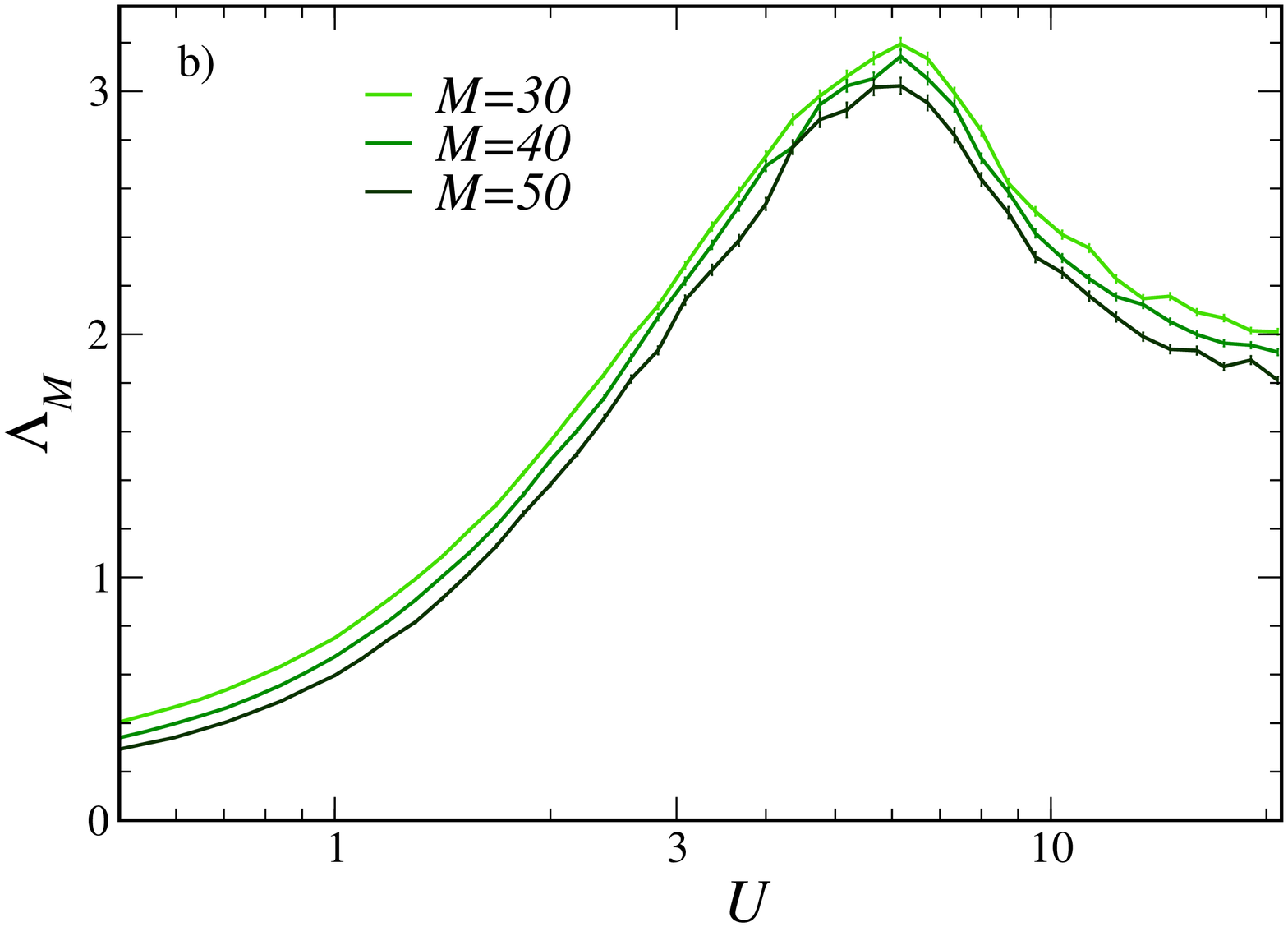}
 	\caption{Reduced localization length of the pair plotted as a function of the interaction strength for increasing values of the transverse size $M$ of the grid. The disorder strength is fixed to $W=9$ and the pair has zero total energy, $E=0$. In Panel (a) the curves for $M=\{8, 10, 12, 16, 20\}$ and $L=400$ are plotted, averaging the Lyapunov exponent over $N_{tr}=\{600, 600, 600, 1000\}$ configurations of the random potential respectively.  
	In Panel (b) the same quantities are shown for larger grids with transverse sizes $M=\{30, 40, 50\}$ and $L=500$. The corresponding number of different disorder realizations used is $N_{tr}=\{3600,4400,2850\}$, respectively.
 	}
 	\label{Fig:intro}
 \end{figure}
As we notice from Fig. \ref{fig:Lambda2d}, $\Lambda_{M}$ possesses a nonmonotonic dependence on $U$, as previously found in one dimension~\cite{Frahm:EigStructAL1DTIP16}, with a peak near $U=6$ whereas, at the band centre ($E=0$), $\Lambda_{M}$ presents the symmetry $\Lambda_{M}(-U)=\Lambda_{M}(U)$. As shown in Panel (a) of Fig. \ref{fig:Lambda2d}, the curves corresponding to different values of $M$ between $M=8$ and $M=20$ intersect each other around $U=1$, suggesting a possible phase transition, as already reported in Refs.~\cite{Ortugno:AndLocTIPDeloc:EPL99,Roemer1999}. 

This result clearly conflicts with the general expectation that 2D Anderson transitions are forbidden in models where both time-reversal and
spin-rotational symmetries are preserved. For the effective model $K$, the two invariances are inherited from the single-particle Anderson model.
In order to test the stability of the result against finite-size effects, we have performed additional calculation using larger grids, going from $M=30$ to $M=50$. 
The results, plotted in Panel (b) of Fig.\ref{fig:Lambda2d} show that the crossing points have completely disappeared, suggesting that the pair localizes in an infinite lattice,
 irrespectively of the specific value of $U$. We therefore conclude that the former results ~\cite{Ortugno:AndLocTIPDeloc:EPL99,Roemer1999} were plagued by severe finite-size effects, due to the limited system sizes accessible at that time, and no Anderson transition can take place for a pair subject to short-range interactions in 
 2D disordered lattices. Our work therefore reconciles the numerics with the one-parameter scaling theory of localization and the concept of universality classes.
 
Based on a least-square variational procedure~\cite{McKinnonKramer:TransferMatrix:ZPB83}, in Ref.~\cite{Stellin:No2BATs2D:2020} we have verified that our numerical results for the reduced localization length obey the one-parameter
scaling ansatz $\Lambda_M=f(\xi/M)$, where $f(x)$ is the scaling function and $\xi$ is  the localization length of the infinite system, up to a scale factor. Notice that the scale factor
is a pure number, independent of the value of the interaction strength. For weak to moderate interactions,  $\xi$ displays an exponential increase over more than 
three orders of magnitude, reaching its maximum value around $U\sim 6$. 

We have also investigated the case of a non-zero energy of the pair. In particular  our numerical results show that $\Lambda_{M} (E,U,W) \leq \Lambda_{M}(0,U,W)$. As a consequence, the pair localizes with an even shorter localization length with respect to the $E=0$ case~\cite{Stellin:No2BATs2D:2020}.

\section{Two-body mobility edges in 3D}

 \begin{figure}
           \label{fig:PhaseDiagE0EU}
 	\includegraphics[width=0.485\textwidth]{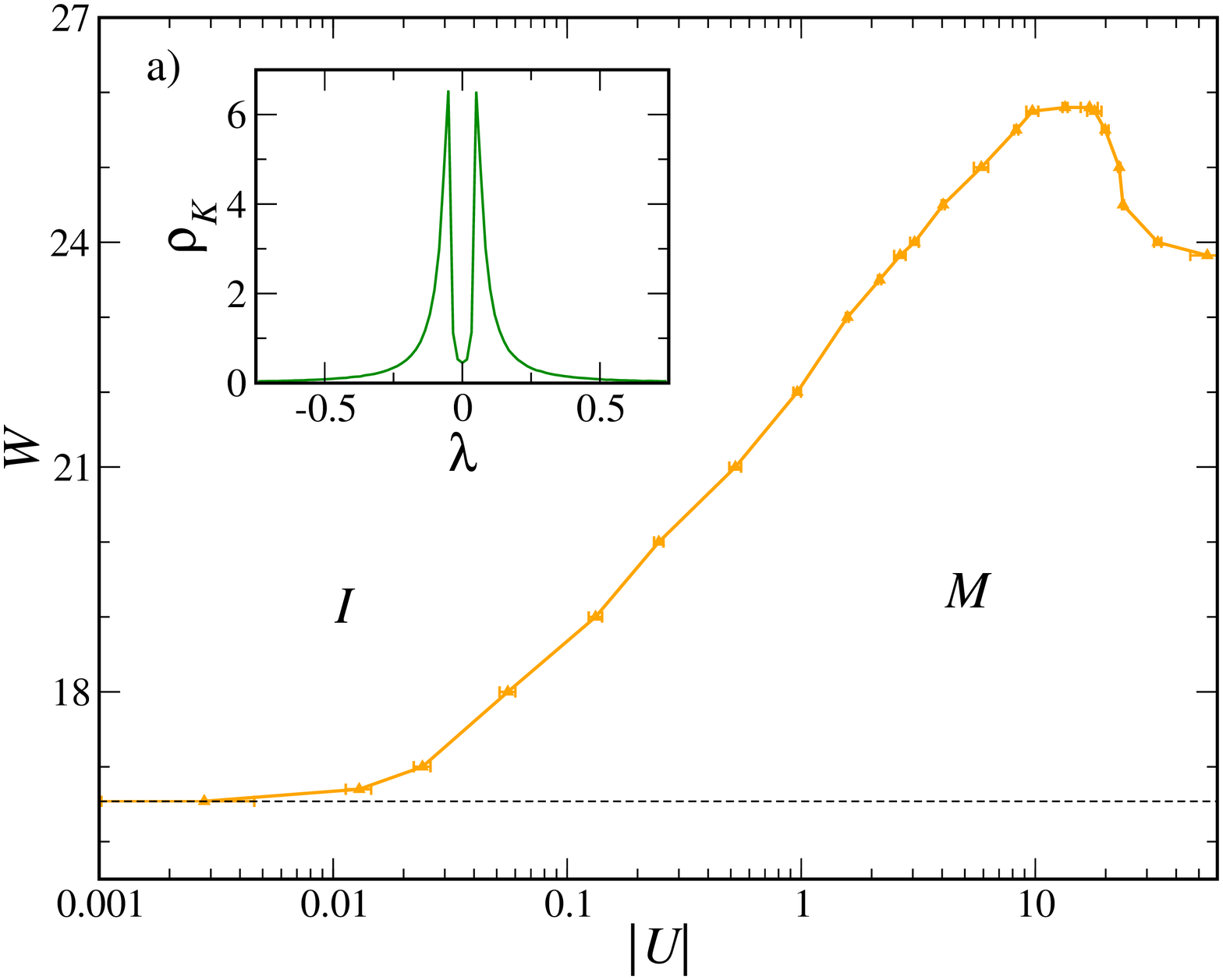}
 	\hspace{\fill}
 	\includegraphics[width=0.485\textwidth]{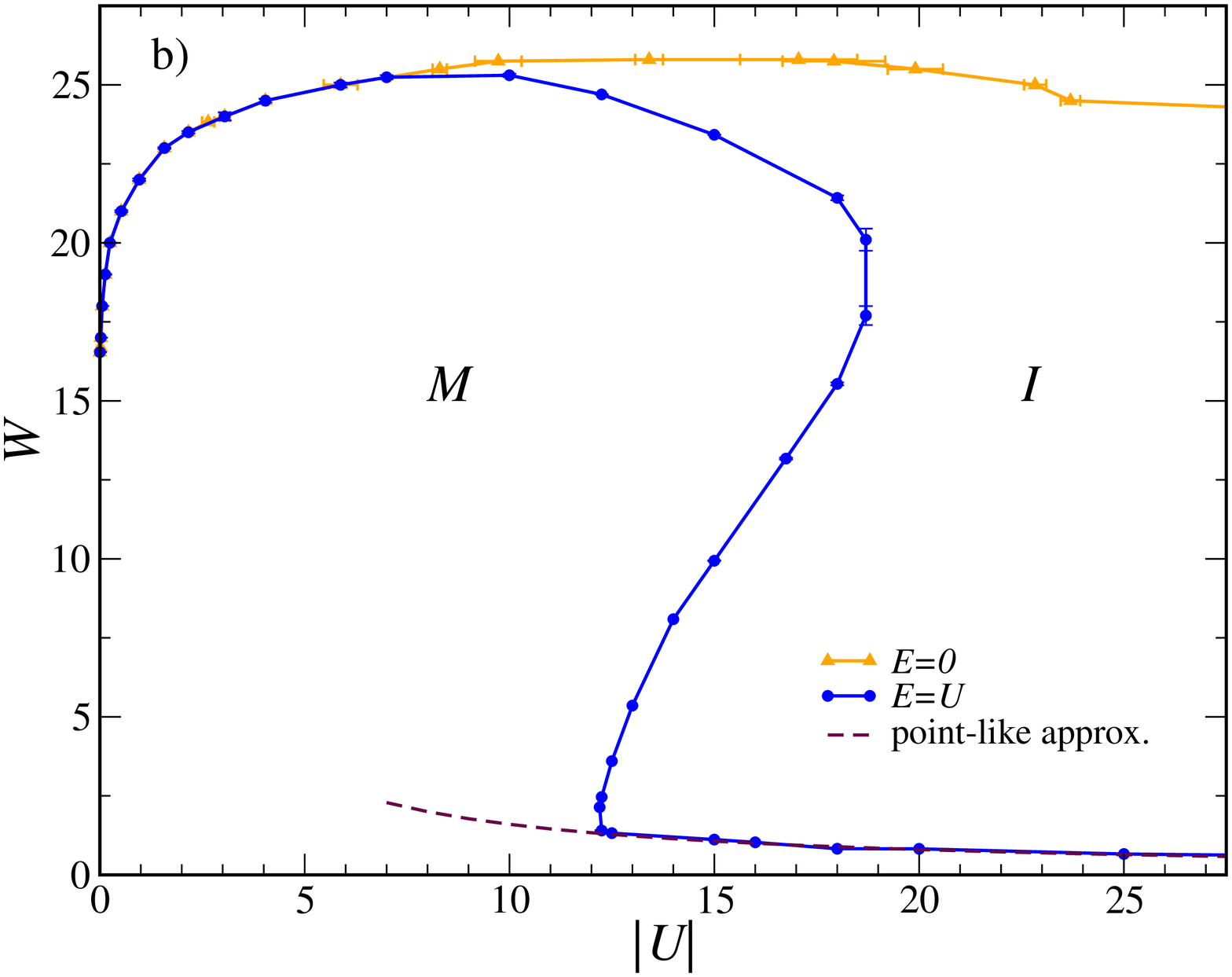}
 	\caption{(a) Phase boundary between localized (labelled by $I$) and extended ($M$) states  in the $(U,W)$ plane for a pair
with zero total energy, $E=0$. The dashed horizontal  line corresponds to the noninteracting limit, $W=W_c^{sp}=16.54$. \emph {Inset:} disorder-averaged density of states $\rho_K$ of the effective Hamiltonian of the pair calculated for $W=23.5$ using a cubic grid of sizes $L=M=20$ with periodic boundary conditions. (b) Phase diagram  for a pair with total energy $E=U$ (blue circles data)~\cite{Stellin:3DPhaseDiags:2020}. The orange triangles data refer to the phase boundary at $E=0$, calculated in Ref.~\cite{Stellin:PRB2019}, which is added for comparison.
The dot-dashed purple line at low disorder corresponds to the molecular result, $W_c\simeq 16.0/|U|$, obtained by treating the pair as a point-like molecule obeying an effective Anderson model, see Eq.~\eqref{eqn:EffTBAndersonMod}. The two diagrams hold for both attractive and repulsive interactions.
  }
 \end{figure}
The scenario changes when three-dimensional systems are examined. We have first determined the critical point for zero total energy of the pair,
$E=0$, and fixed disorder strength $W=23.5$. By carrying out the same analysis presented in the previous section, we identified a critical region where the curves $\Lambda_{M}(U)$ obtained for different $M$ cross each other. Since the values of $M$ numerically accessible are rather small, the crossing point drifts to stronger interaction and upwards as the
system sizes increase. For the largest system sizes, the height of the crossing point becomes close to the universal value, $\Lambda_{c,\mathrm{orth}}=0.58\pm0.09$, expected for the orthogonal universality class, within the accuracy of our numerics. Then we use this information to pinpoint the position of the critical point by including the leading irrelevant variable
$y_{orth}=3.3\pm 0.6$~\cite{Slevin:CriticalExponent:NJP14} in the scaling analysis, thus accounting for the drift of the crossing point. In this way we were able to identify the critical interaction strength as $U_c=2.16\pm 0.04$~\cite{Stellin:PRB2019}. 

The same procedure has then been systematically applied to explore the phase diagram in the space of energy, disorder and interaction. 
We first determined the phase boundary at zero total energy of the pair, which is displayed in Fig. \ref{fig:PhaseDiagE0EU}.  We see that, for zero total energy of the pair, 
Anderson transitions occur for disorder strength $W>16.5$, where all single-particle states are localized~\cite{Slevin:CriticalExponent:NJP14}.  The critical disorder strength 
exhibits a nonmonotonic behaviour as a function of the interaction, reaching its maximum value at $U\sim 14$. This feature can be understood from the behaviour of the disorder-averaged density of states $\rho_K(\lambda)=\overline{\sum_r \delta(\lambda-\lambda_r)/N}$ of the effective model, which is plotted in the inset of Panel (a) of Fig. \ref{fig:PhaseDiagE0EU}. Notice that for $E=0$ this quantity is 
  parity-symmetric, ($\rho_K(\lambda) = \rho_K(-\lambda)$), like the reduced localization length.
The weakly interacting states, located in the tails of the density of states, where $\rho_K \propto \lambda^{-2}$, localize at lower disorder than those close to the peaks, occurring at $\lambda = \pm W^{-1}$. 

Making $J$ explicit, in the strongly interacting regime, that corresponds to $E\sim U$ with $|E|\gg W,J$, the effective Hamiltonian takes the same form of the single-particle Anderson model, 
but endowed with modified parameters, describing the motion of  tightly bound pairs~\cite{Dufour:PRL2012,Stellin:3DPhaseDiags:2020}
\begin{equation}
\hat{K}_{\mathrm{TB}}\Psi(\pmb{n})=\frac{2J^{2}}{E}\sum\limits_{i=1}^{3}\Psi(\pmb{n}+\pmb{\mathrm{e}}_{i})+\frac{2J^{2}}{E}\sum\limits_{i=1}^{3}\Psi(\pmb{n}-\pmb{\mathrm{e}}_{i})+\frac{2v_{\pmb{n}}(E+2v_{\pmb{n}})}{E}\Psi(\pmb{n}) =E_{\mathrm{TB}}\Psi(\pmb{n})\hspace{0.1cm},
\label{eqn:EffTBAndersonMod}
\end{equation}
where the tunnelling amplitude is $J_{\mathrm{TB}}=-2J^{2}/E$, the disorder typical value is $W_{\mathrm{TB}}=2W$, and $E_{\mathrm{TB}}=E^2(\lambda-1/E-12J^2/E^3)$ denotes the effective energy of the molecule. Notice that $|J_{\mathrm{TB}}|\sim 2J^2/|U| \gg J $, implying that the molecule localizes already for weak disorder. In particular we can estimate the position of the mobility edge for tightly bound molecules from the corresponding results for the single-particle Anderson model, upon substitution of the bare parameters with the molecular ones.

 \begin{figure}
 	\includegraphics[width=0.333\textwidth]{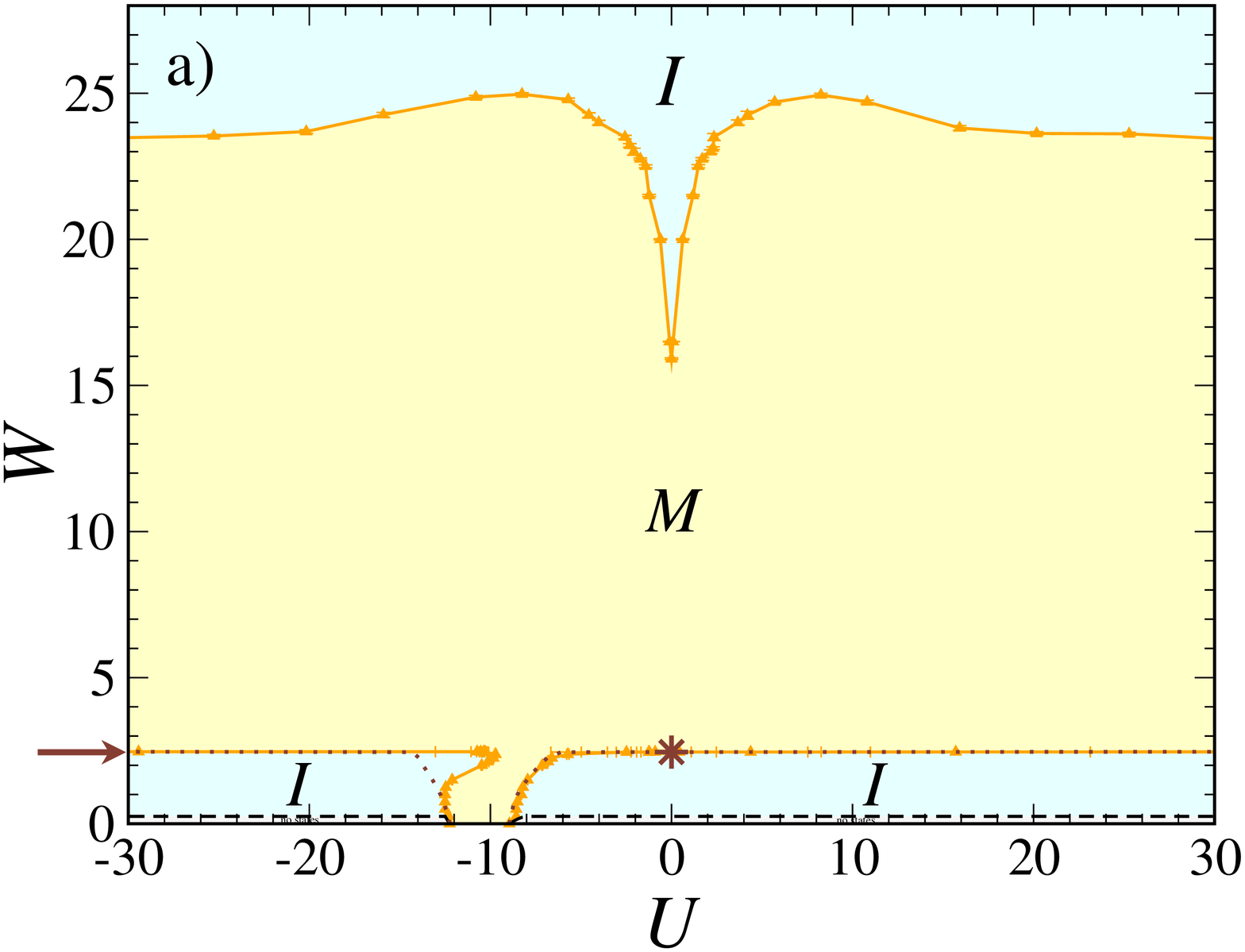}
 	\includegraphics[width=0.333\textwidth]{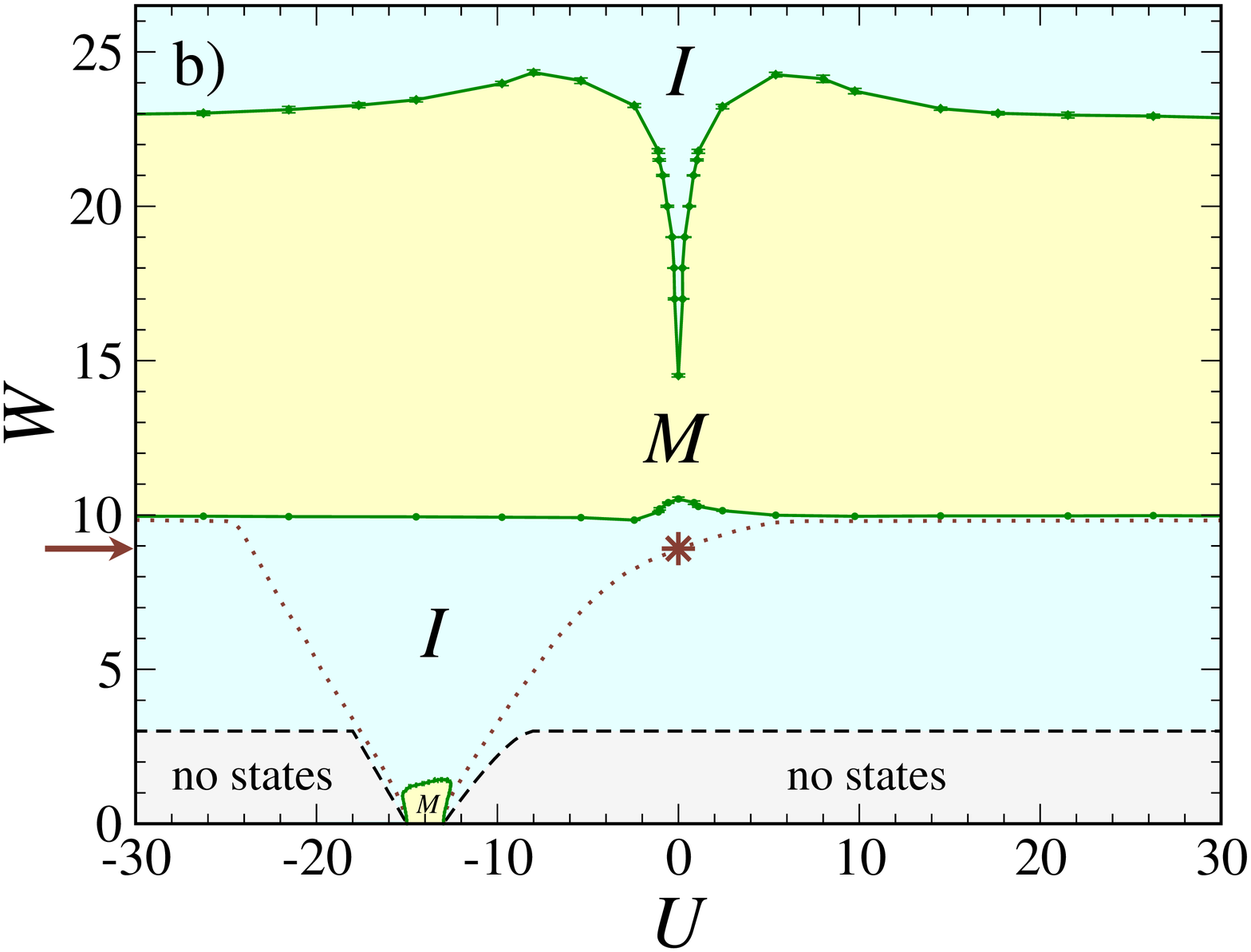}
 	\includegraphics[width=0.333\textwidth]{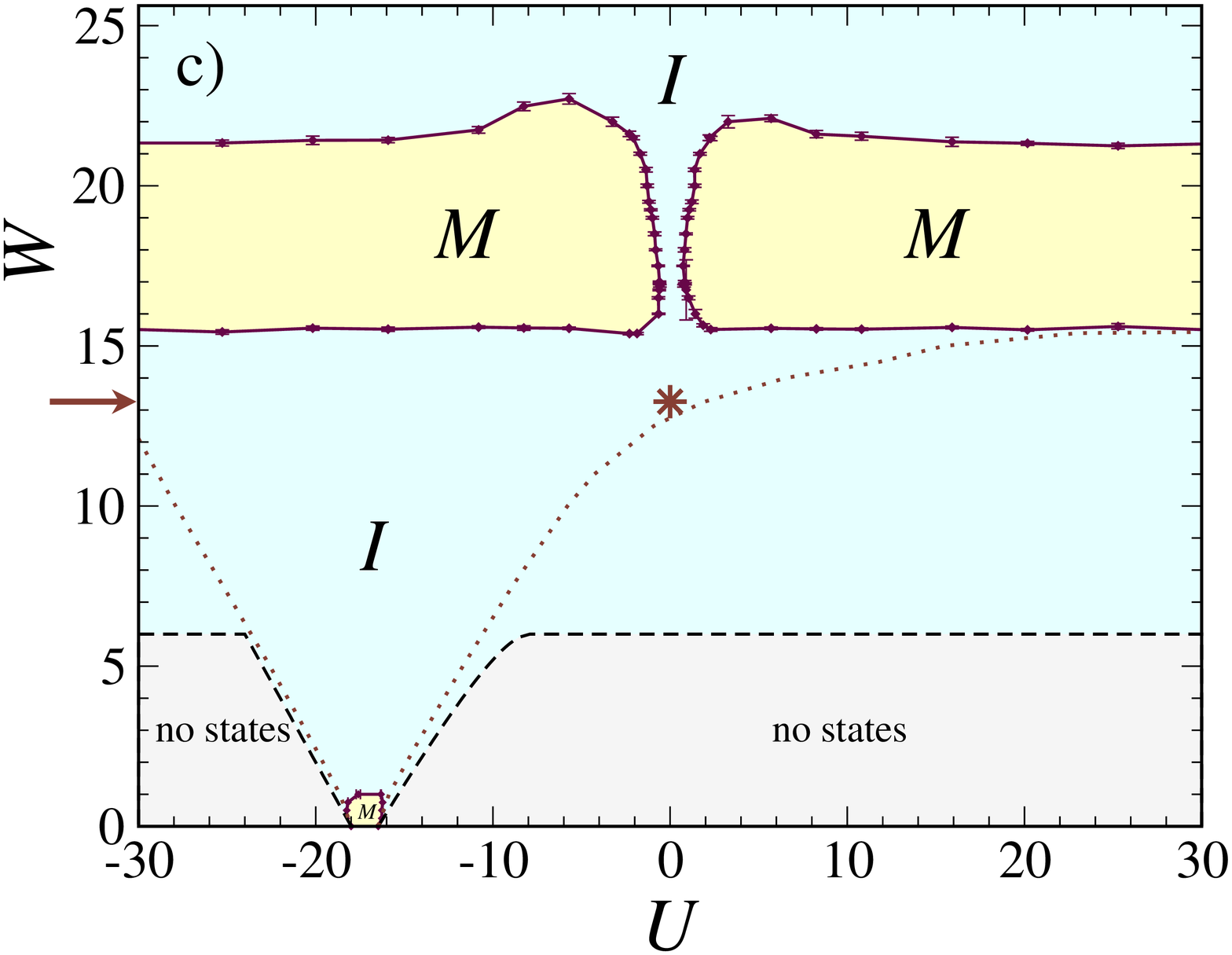}
 	\hspace{\fill}
 	\caption{Phase diagrams in the $(U,W)$ plane for a pair with $E=-12.25$ (a), $E=-15$ (b) and $E=-18$ (c). The phase boundaries are denoted by the orange, green and purple triangles respectively. The numerical band edges are represented by the dotted lines, whereas the black dashed lines denote the exact band edges, which delimit the regions in which no two-particle states are present, shaded in light grey. The brown arrows indicate the disorder strength at which the molecular state evolve into a scattering state. For this value of disorder the right numerical band edge crosses the $U=0$ axis, as signalled by the brown stars.
 	}
 	\label{fig:3dFixedEPhaseDiags}
 \end{figure}
 
In the Panel (b) of Fig. \ref{fig:PhaseDiagE0EU} we display the phase boundary between localized and extended states calculated along the direction 
$E=U$. For weak interactions, the mobility edge remains very close to the result previously obtained for $E=0$. In the opposite strongly interacting regime,
 the system exhibits a single Anderson transition, corresponding to the localization of tightly bound molecules with $W_c\simeq 16/|U|$, as evidenced by the purple dot-dashed curve
 in Fig. \ref{fig:PhaseDiagE0EU} (b). For intermediate values of the interaction strength, $12 \lesssim U \lesssim 19$, 
 the critical disorder strength displays an interesting s-like shape, signalling that
 the system undergoes three different metal-insulator transitions (instead of one)  at given interaction strength. 
 
To the purpose of understanding the emergence of this region, we have built
 the three diagrams at fixed energy displayed in Fig. \ref{fig:3dFixedEPhaseDiags}, corresponding to $E=-12.25, -15, -18$. We notice that, at very low disorder, the pair exhibits
 a metallic phase corresponding to delocalized molecule, which shrinks in size as $E$ becomes large and negative, due to the increased inertia of the bound state.
 A striking feature of the three phase diagrams is the existence of disorder-induced delocalization transitions, resulting in a metallic  phase for intermediate values of $W$.
This counterintuitive result reflects a fundamental change in the nature of the pair state as the disorder strength increases.
For sufficiently weak disorder and $|E|>12$, the pair behaves as a molecule and, in particular, its energy falls outside the noninteracting two-particles energy spectrum. Neglecting 
Lifshitz tails~\cite{Lifshitz:IDoSTails:1963}, the latter corresponds to the interval $[-2\varepsilon_{\mathrm{BE}},2\varepsilon_{\mathrm{BE}} ]$, where $\pm \varepsilon_{\mathrm{BE}}(W)$
are the numerical band edges of the single-particle model, which can be easily determined by making use of the coherent potential approximation (CPA)~\cite{Elliott:CPApprox:74,Vollhardt:SelfConsistentTheoryAnderson:92}. As $W$ increases, the noninteracting energy spectrum broadens and the energy $E$ will inevitably fall inside it, transforming 
the molecule into a scattering state, which is more akin to delocalize. This change of behaviour is highlighted by the brown lateral arrows in the phase diagrams of Fig. \ref{fig:3dFixedEPhaseDiags}. 
  
In Fig. \ref{fig:3dFixedEPhaseDiags}, the phase boundary for scattering states (at high disorder) does not depend on the sign of the interaction, and, when the latter vanishes, the critical disorder is consistent with the one found in the single-particle case~\cite{Bulka:ZPB1987}. In accordance with Ioffe-Regel criterion~\cite{Ioffe-Regel:Criterium:60}, the  states in the limit of infinite interactions (in magnitude) must possess a large mean free path, since the phase boundary follows closely the numerical bande edge, which is represented by the dotted brown curve and delimits the region where $\rho_{K}$ is negligible. The nonmonotonicity of the upper phase boundary of extended scattering states is clearly reminiscent of the one computed for $E=0$ (see Fig. \ref{fig:PhaseDiagE0EU} (a)), but the maxima of $W_c$ generally occur for lower values of the disorder. 

From the three panels of Fig. \ref{fig:3dFixedEPhaseDiags}, it is possible to observe that the single metallic region for $E=-12.25$ (panel a) splits into two regions for $E=-15$ related to molecular and scattering states (panel b), then the latter break up in turn into two disconnected regions for $E=-18$ (panel c). In that case, there are no delocalized states at vanishing interactions, in accordance with the single-particle case~\cite{Bulka:ZPB1987}.  

Finally, the grey regions of the diagrams  of Fig. \ref{fig:3dFixedEPhaseDiags} correspond to configurations unaccessible to the system, where the density of state is
exactly zero. These regions are limited by the  black  dashed curves, corresponding to the rigorous band edges, which have been calculated analytically~\cite{Stellin:3DPhaseDiags:2020}. 

\section{Conclusions}

Based on large-scale numerics, we have investigated the localization properties of a two-particle system described by the Anderson-Hubbard model.
We have first proved that Anderson transitions for the pair do not occur in two dimensions, as opposed to previous claims~\cite{Ortugno:AndLocTIPDeloc:EPL99,Roemer1999}. In three dimensions, we have also shown that the universality class of the transition is not affected by interactions and that the critical disorder at the band centre is a nonmonotonic function of interaction strength, a feature we have interpreted by analyzing the behaviour of the disorder-averaged density of states. 
The phase diagram gets particularly rich when the total energy is finite and becomes larger (in modulus) than the bandwidth, corresponding to $|E|>12$.
In this case, the phase diagram exhibits multiple Anderson transitions as the disorder increases, reflecting the underlying change from molecular to scattering states.  Our results represent the zero-particle-density limit of the many-body localization problem~\cite{Altshuler:MetalInsulator:ANP06,Abanin:RMP2019}.

\section{Acknowledgements}

We acknowledge D. Basko, N. Cherroret, D. Delande, K. Frahm, C. Monthus, S. Pilati, T. Roscilde and S. Skipetrov for fruitful discussions.
This project has received funding from the European Union's Horizon 2020 research and innovation programme under the Marie Sklodowska-Curie grant agreement No 665850. 
The numerical study of the two-body localization problem reviewed in this work has required in total about three millions hours of CPU times.
For its realization, access was granted to the HPC resources of CINES (Centre Informatique National de l'Enseignement Sup\' erieur) under the allocations 2016-c2016057629, 2017-A0020507629, 2018-A0040507629, 2019-A0060507629 and 2020-A0080507629 supplied by GENCI (Grand Équipement National de Calcul Intensif).\\

\section{References}

\bibliographystyle{iopart-num}
\bibliography{Biblio2BodyProc}

\end{document}